\documentstyle[psfig]{mn}
\footnotesize
\newdimen\digitwidth    
\setbox0=\hbox{\rm0}
\digitwidth=\wd0
\catcode`!=\active
\def!{\kern\digitwidth}
\normalsize

\title{Pulsar rotation measures and the magnetic structure of our Galaxy}

\author[J. L. Han et al.]
       {
        J. L. Han$^{1,2}$,
        R. N. Manchester$^3$,
	and
        G. J. Qiao$^{2,4,5}$
\\
$^1$Beijing Astronomical Observatory, Chinese Academy of Sciences (CAS),
        Beijing 100080, China \\
$^2$Beijing Astrophysical Center, CAS-PKU, Beijing 100871, China\\
$^3$Australia Telescope National Facility, CSIRO, PO Box 76,
        Epping, NSW 2121, Australia \\
$^4$CCAT (World Laboratory), PO Box 8730, Beijing 100080, China \\
$^5$Department of Geophysics, Peking University (PKU), Beijing 100871, China
	}

\date{Accepted  Feb.3, 1999;
        Received Jan.20, 1999;
	in original form 1997 November 6
     }

\pubyear{1999}

\begin{document}

\maketitle


\begin{abstract} We have obtained 63 rotation measures (RMs) from
polarization observations of southern pulsars, of which 54 are new
measurements and 3 are varied from previous values.
The new pulsar RM data at high Galactic latitudes are mostly
consistent with the antisymmetric RM distribution found previously. For the
Galactic disc, evidence for a field reversal near the Perseus arm, and
possibly another beyond it, is presented. Inside the Solar Circle, in
addition to the two known field reversals in or near the Carina-Sagittartus
arm and the Crux-Scutum arm, a further reversal in the Norma arm is
tentatively identified. These reversals, together with the pitch angle
derived from pulsar RM and stellar polarization distributions, are
consistent with bisymmetric spiral (BSS) models for the large-scale
magnetic field structure in the disc of our Galaxy. However, discrimination
between models is complicated by the presence of smaller-scale
irregularities in the magnetic field, as well as uncertainties in the
theoretical modelling.
\end{abstract}


\section{Introduction}

Early observations of starlight polarization (eg. Mathewson \& Ford 1970),
the rotation measure (RM) distribution of extragalactic radio sources
(e.g. Gardner, Morris \& Whiteoak 1969; Simard-Normandin \& Kronberg 1980)
and pulsar RMs (Manchester 1974) revealed many of the principal features
of the Galactic magnetic field. See Han \& Qiao (1994) for a
comprehensive review of earlier work.

Pulsars have a number of advantages as probes of the interstellar magnetic
field. They typically have substantial linear polarization, making the
Faraday rotation relatively easy to measure; they are distributed in the
Galaxy with approximately known distances, allowing the three-dimensional
properties of the magnetic field to be investigated; and they apparently
have no intrinsic rotation measure (Manchester 1971). With the dispersion
measure (DM), direct estimates of the Galactic magnetic field strength,
averaged over the line of sight to the pulsar with a weight proportional to
the local free electron density, can be obtained from 
\begin{equation}
\langle B_{||} \rangle =1.232 \; {\rm RM/DM} 
\end{equation} 
where $B_{||}$ is in $\mu$G and RM and DM are in their usual units, rad
m$^{-2}$ and cm$^{-3}$ pc, respectively.  For a positive RM, the mean field
is directed toward us.  Manchester (1974) used RMs of 38 pulsars to estimate
the strength and direction of the local Galactic magnetic field, concluding
that the local field is basically azimuthal. Thomson \& Nelson (1980)
analysed the RMs of 48 pulsars, finding evidence for a field reversal in the
inner Carina-Sagittarius spiral arm.

More extensive studies became possible only after the RMs of
a large sample of pulsars had been measured. Lyne \& Smith (1989) analysed
the RMs of 185 pulsars (Hamilton \& Lyne 1987), and found evidence for a
reversal about 1 kpc inside the Solar Circle. Comparison of pulsar and
extragalactic RMs  in the anticentre direction indicated a further reversal
at larger Galactocentric radius. Rand \& Lyne (1994) added more pulsar
RMs toward the inner Galactic Plane, and found evidence for
two reversals, one about 0.4 kpc inside the Solar Circle, between the Sun
and the Sagittarius arm, and another at a Galactocentric radius of 5.5 kpc
(with the Sun at 8.5 kpc), near or within the Scutum arm.

At least three classes of model have been proposed for the global structure
of magnetic field in the disc of our Galaxy: concentric rings, an asymmetric
spiral (ASS) or a bisymmetric spiral (BSS).  Rand \& Kulkarni (1989) and
Rand \& Lyne (1994) argued that pulsar RMs are consistent with a
concentric-ring model for the field. Vall\'ee (1991, 1996) has proposed that
the field has an ASS form. Although a pure ASS structure has no reversals,
modelling suggests that, depending on the initial conditions, reversals can
be present (Poezd et al. 1993) and Vall\'ee has argued
that the presence of two field reversals inside the Solar Circle is
consistent with the ASS model.  Early analyses of the RM distribution of
extragalactic radio sources (Simard-Normandin \& Kronberg 1980; Sofue \&
Fujimoto 1983) suggested that the Galactic magnetic field has a BSS form, in
which the field direction reverses from arm to arm (Heiles 1995,1996ab).
Such structures have
been deduced from radio polarization observations of nearby spiral galaxies,
such as M81 (e.g.  Krause, Beck \& Hummel 1989). A statistical study of RMs
of a selected sample of pulsars at distances less than $\sim 3$ kpc (Han \&
Qiao 1994, 1996) also supported the BSS model for our Galaxy. The pitch
angle
\footnote{There is much confusion in the literature about the definition
of pitch angle. For our Galaxy, various authors (e.g.  Georgelin \&
Georgelin 1976; Simard-Normandin \& Kronberg 1980; Dame et al. 1986;
Grabelsky et al. 1988; Amaral \& Lepine 1997) have defined pitch angles
to be positive, whereas about an equal number (e.g. Sofue \& Fujimoto 1983;
Rand \& Kulkarni 1989; Han \& Qiao 1994; Vall\'ee 1995; Beck et al. 1996;
Heiles 1996a) have defined them to be negative.

We propose the following {\it physical} definition, in principle applicable to
all galaxies: the galactic azimuthal angle $\phi$ is {\it defined}
to be increasing in the direction of galactic rotation. Logarithmic spirals
are then defined by $$ R = R_0 \exp(k\phi),$$ where $R$ is the radial
distance and $R_0$ is a scale radius. The pitch angle is $p=\arctan(k)$.
This is {\it negative} for trailing spirals such as our Galaxy, where $R$
increases with decreasing azimuthal angle $\phi$. For our Galaxy, the
Galactic angular momentum vector points toward the {\it south} Galactic
pole, and $\phi$ increases in a clockwise direction when viewed from the
north Galactic pole.}
of this large-scale field is about $-8\degr$.

The magnetic field in the thick disc or halo of our Galaxy may have a
different origin to the disc field. Han et al. (1997) showed that the
RM distributions of extragalactic radio sources and pulsars in the inner
Galaxy (between $l=270\degr$ and $90\degr$) are antisymmetric with respect
to the Galactic Plane and the meridian through the Galactic Centre.
This antisymmetry is not caused by local perturbations, for example,
by Loop I, but results from large-scale toroidal magnetic field structure
in the thick disc or halo, which has opposite directions above and below
the Galactic plane, as in an A0 dynamo mode (see Fig. 19a of Wielebinski
\& Krause 1993).

Clearly, further work is needed to determine the field structure of our
Galaxy, especially that in the disc, and to clarify whether or not the
fields in the disc and in the halo are generated or maintained by different
mechanisms and are independent of each other (Sokoloff \& Shukurov 1990).
Observations of pulsar RMs provide a good opportunity to investigate these
issues further. There are about 750 known pulsars but, up to now, only 262
pulsars have known RMs (see updated catalogue of Taylor, Manchester \& Lyne
1993; TML93). In any quantitative analysis of the disc field, the RMs of
high-latitude pulsars should be excluded, further limiting the usable
pulsar sample. Furthermore, the paucity of measured RMs in the third and
fourth Galactic quadrants causes significant uncertainties in the analysis
(Han \& Qiao 1994).

Between 1996 August 31 and September 5, polarization observations of 66
southern pulsars were obtained with the ATNF Parkes 64-m telescope, mainly
at 436 and 660 MHz (Manchester, Han \& Qiao 1998).  Most of these pulsars
were discovered in the recent Parkes southern pulsar survey (Manchester et
al.  1996; Lyne et al. 1998). In Sect. 2 of this paper we present RMs for
most of these pulsars. We analyse the high-latitude magnetic features of our
Galaxy in Sect. 3 and the magnetic field structure in the thin disc in
Sect. 4. Current models are discussed in the light of the observational
evidence in Sect. 5, and conclusions are presented in Sect. 6.

\begin{figure}		
\psfig{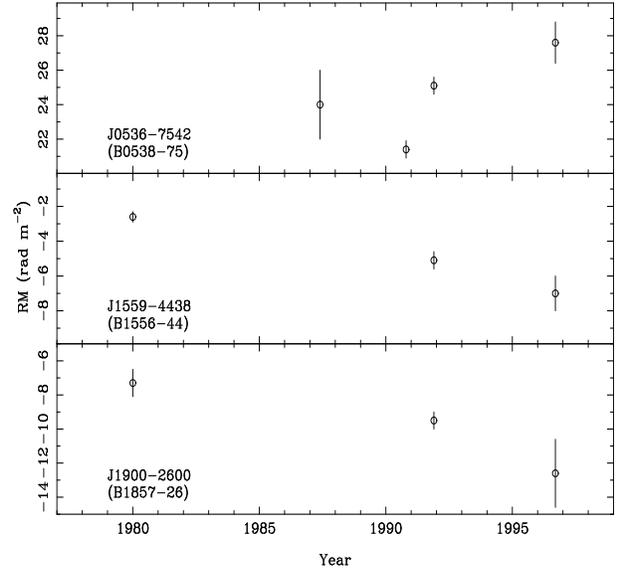}
\caption[]{RM variations of three pulsars. Data of 1980 are taken from
Taylor et al. (1993), unpublished data of Hamilton,
McCulloch \& Manchester (1981; HMM81), of 1987 from Costa et al.  (1991), of
1990 from Qiao et al. (1995), of 1991 from van Ommen et al. (1997), and of
1996 from this paper.}
\end{figure}

\section{Pulsar Rotation Measures}

\begin{table*}
\begin{minipage}{175mm}
\caption{Rotation Measures for 63 pulsars}
\begin{small}
\begin{tabular}{lrrrrrrclrrrrrr}
\hline
\multicolumn{1}{c}{PSR J}& 
\multicolumn{1}{c}{$l$} &
\multicolumn{1}{c}{$b$} &  Dist &  DM$^*$ & RM  & Err & 
\hspace{1mm} &
\multicolumn{1}{c}{PSR J}& 
\multicolumn{1}{c}{$l$} &
\multicolumn{1}{c}{$b$} &  Dist &  DM$^*$ & RM  & Err
 \\
 & ($\degr$)~~ & ($\degr$)~~ & (kpc) & 
& \multicolumn{2}{c}{(rad m$^{-2}$)} & &
 & ($\degr$)~~ & ($\degr$)~~ & (kpc) & 
& \multicolumn{2}{c}{(rad m$^{-2}$)} \\
\hline
0108$-$1431 &140.9&$-$76.8&   0.10&  2.4&  $-$1  &  3  & &
1544$-$5308& 327.3&    1.3&   1.29& 35.2& $-$29  &  7  \\
0133$-$6957 &297.7&$-$46.7&$>$2.42& 23.0&     7  &  1  & &
1549$-$4848& 330.5&    4.3&   1.55& 57.0& $-$15  &  8  \\ 
0134$-$2937 &230.3&$-$80.3&$>$1.78& 21.8&    13  &  2  & & 
1557$-$4258 &335.3&    8.0&   7.54&144.5& $-$37  &  2  \\
0255$-$5304 &269.9&$-$55.3&   1.15& 15.9& $-$10  & 12  & &
1559$-$4438$^d$& 334.5&    6.4&   1.63& 58.8&$-$7&  1  \\ 
0459$-$0210 &201.4&$-$25.7&   1.30& 21.0&    18  &  9  & & 
1559$-$5545& 327.2& $-$2.0&   5.04&211.0&$-$150  & 20  \\[2mm]
0520$-$2553 &228.4&$-$30.5&$>$3.47& 33.8&    19  & 15  & &
1603$-$5657& 326.9& $-$3.3&   8.52&264.1&    27  &  5  \\ 
0536$-$7543$^a$ &287.2&$-$30.8&   1.05& 17.6&  28&  2  & & 
1604$-$4909& 332.2&    2.4&   3.59&140.8& $-$16  &  6  \\
0540$-$7152 &282.1&$-$31.2&   2.69& 29.4&    43  & 15  & & 
1604$-$7203 &316.7&$-$14.6&   2.56& 54.4&    22  & 11  \\ 
0742$-$2822$^b$ &243.8& $-$2.4&   1.90& 73.8& 156&  5  & & 
1622$-$4332& 338.3&    4.3&   8.07&230.7&   140  & 20  \\
0856$-$6137 &278.6&$-$10.4&   6.52& 95.0& $-$70  & 18  & &
1648$-$3256& 349.6&    7.8&   5.30&128.3& $-$60  & 15  \\[2mm] 
0857$-$4424 &265.5&    0.8&   6.51&184.4& $-$75  & 20  & &
1651$-$5222& 335.0& $-$5.2&   6.39&179.1& $-$38  &  5  \\
0924$-$5302 &274.7& $-$1.9&   5.74&155.1&   150  & 20  & &
1700$-$3312& 351.1&    5.5&   5.62&166.8& $-$15  &  3  \\   
0934$-$5249& 275.7& $-$0.7&   2.90& 99.4&    18  &  6  & &
1722$-$3712& 350.5& $-$0.5&   2.51& 99.5&   104  &  3  \\
0942$-$5657& 279.3& $-$3.0&   5.07&160.8&   135  &  4  & &
1741$-$3927$^e$& 350.6& $-$4.7&   4.75&158.5& 180& 14  \\ 
0955$-$5304& 278.3&    1.2&   4.86&156.9& $-$97  & 13  & & 
1750$-$3503& 355.3& $-$4.1&   5.28&189.4&   173  & 11  \\[2mm]
1034$-$3224 &272.1&   22.1&$>$4.68& 50.9&  $-$8  &  1  & &
1751$-$4657$^f$ &345.0&$-$10.2&   1.03& 20.3& 19 &  1  \\ 
1036$-$4926& 281.5&    7.7&   8.39&135.0& $-$11  &  6  & &
1801$-$0357 & 23.6&    9.3&   6.66&117.6&    32  & 11  \\
1042$-$5521 &285.2&    3.0&   6.95&306.0&   155  &  5  & &
1808$-$0813&  20.6&    5.8&   5.19&151.0&    73  &  7  \\
1116$-$4122$^c$ &284.5&   18.1&   2.77& 40.5&  31& 16  & &
1829$-$1751$^g$&  14.6& $-$3.4&   5.52&217.8& 306&  6  \\
1123$-$4844 &288.3&   11.6&$>$8.74& 92.9&  $-$7  &  5  & &
1849$-$0636&  26.8& $-$2.5&   3.66&147.6& $-$35  & 21  \\[2mm]
1126$-$6942 &295.6& $-$8.1&   2.07& 55.3& $-$31  &  4  & &
1852$-$2610 &  9.5&$-$11.9&   2.26& 56.8& $-$21  &  1  \\
1202$-$5820& 296.5&    3.9&   4.96&145.8&   139  &  3  & & 
1900$-$2600$^h$ & 10.3&$-$13.5&   1.70& 38.1&$-$13& 2  \\
1210$-$5559& 297.1&    6.5&  11.83&174.0&    58  &  1  & &
1901$-$0906&  26.0& $-$6.4&   2.42& 73.0&    44  & 12  \\
1253$-$5820 &303.2&    4.6&   2.97&101.0&    31  &  5  & & 
1932$-$3655 &  2.1&$-$23.6&$>$4.41& 59.9&     6  &  3  \\
1328$-$4921 &309.1&   13.1&$>$7.78&118.0&   170  & 20  & &
1946$-$2913 & 11.1&$-$24.1&$>$4.30& 44.2&     8  &  7  \\[2mm]
1328$-$4357 &309.9&   18.4&   2.23& 41.4& $-$41  &  3  & &
1949$-$2524 & 15.2&$-$23.4&   1.32& 22.8& $-$13  &  8  \\
1359$-$6038& 311.2&    1.1&   5.91&295.0&    33  &  5  & &
2038$-$3816 &  3.9&$-$36.7&$>$2.94& 33.9&    30  & 15  \\
1418$-$3921 &320.9&   20.5&$>$5.04& 61.5& $-$15  &  8  & &
2053$-$7200$^i$ &321.9&$-$35.0&   1.01& 16.9& 9  &  4  \\
1440$-$6344& 314.6& $-$3.4&   4.03&124.2&    29  &  4  & &
2108$-$3429 &  9.7&$-$42.2&$>$2.63& 30.2&    50  & 20  \\  
1527$-$5552& 323.6&    0.6&   7.07&358.0&    34  &  4  & &
2144$-$3933 &  2.8&$-$49.5&   0.18&  3.4&  $-$2  & 10  \\[2mm] 
1527$-$3931 &333.1&   14.0&   1.88& 46.8&     4  &  1  & &
2346$-$0609 & 83.8&$-$64.0&$>$1.96& 22.5&  $-$5  &  1  \\
1542$-$5034& 328.6&    3.6&   2.54& 95.0& $-$70  &  8  & & 
            &     &       &       &     &        &     \\
\hline
\end{tabular}
$^*$ DM values in cm$^{-3}$pc.
$^a$PSR B0538$-$75: RM$ = 24 \pm 2 $ by Costa, McCulloch \& Hamilton (1991),
RM$ = 21.4\pm 0.5$ by Qiao et al. (1995), and RM$ = 25.1\pm 0.5$ by van
Ommen et al.  (1997);   
$^b$PSR B0742$-$28: RM$ = 150.4 \pm 0.1$
in TML93 from Hamilton, McCulloch \& Manchester, unpublished data (HMM81); 
%
$^c$PSR B1114$-$41: RM$ = 50 \pm 40 $ by van Ommen et
al. (1997); 
$^d$PSR B1556$-$44: RM$ = -2.6 \pm 0.3 $ in TML93 from HMM81, and
 	RM$ = -5.1 \pm 0.5 $ by van Ommen et al. (1997);	
$^e$PSR B1737$-$39: RM$ = 221 \pm 29 $ by van Ommen et al. (1997); %
%
$^f$PSR B1747$-$46: RM$ =  18\pm 3 $ in TML93 from HMM81 
$^g$PSR B1826$-$17: RM$ = 317\pm16 $ by Hamilton \& Lyne (1987); 
$^h$PSR B1857$-$26: RM$ = -7.3\pm 0.8 $ in TML93 from HMM81, and
	RM$=-9.5\pm0.5$ by van Ommen et al (1997);		
$^i$PSR B2048$-$72: RM$ =  17\pm 1$ by Qiao et al. (1995).	
\end{small}
\end{minipage}
\end{table*}

Pulsar polarization observations were made in bands centred on 436, 660 or
1500 MHz using the Parkes 64-m telescope of the Australia Telescope National
Facility. Dual-channel cryogenic systems receiving orthogonal linear
polarizations were used at each frequency. Signals were processed in the
Caltech correlator (Navarro 1994), which gives 128 lags in each of four
polarization channels and folds data synchronously with the pulsar period in
up to 1024 bins per period. The data were transformed to the frequency
domain, calibrated and dedispersed to form either 8 or 16 frequency
sub-bands and corrected for variations in parallactic angle and ionospheric
Faraday rotation. Further observational details can be found in Manchester
et al. (1998).

In off-line analysis, data from each observation were summed to form two
sets of Stokes-parameter profiles for the upper and lower halves of the
bandpass. RMs were then determined by taking weighted means of
position-angle differences for bins where the uncertainty was less than
$10\degr$; RM uncertainties were computed from the scatter in position angle 
differences.  For highly dispersed and weakly polarized profiles, we searched
the sub-band data in RM for a peak in the linearly polarized intensity $L =
(Q^2 + U^2)^{1/2}$.  If a significant peak was found, the sub-band data were
summed with correction for the Faraday rotation to form the upper and lower
bandpass profiles, and a final value of the RM and its error were determined
from the position-angle differences and uncertainties as described above.

Rotation measures for 63 pulsars determined in this way are listed in Table
1. The values for Galactic longitude and latitude, distance estimate and
dispersion measure in Table 1 are taken from the updated TML93 catalogue.
Previously determined RMs for nine pulsars are indicated by footnotes to the
table. For four of these, PSRs J0742$-$2822, J1116$-$4122, J1741$-$3927,
J1751$-$4657 and J1829$-$1751, the present RMs are either consistent with
the old ones, or more precise. The RM of PSR J2053$-$7200 seems marginally
changed from RM$ = 17\pm1$ rad m$^{-2}$ in 1990 September to RM$= 9\pm4$ rad
m$^{-2}$ in 1996 September. As shown in Fig. 1 the RMs of PSRs J0536$-$7542,
J1559$-$4438 (see also van Ommen et al.  1997) and J1900-2600 have varied
significantly over several years. Similar RM variations were seen in the
Vela (Hamilton et al. 1977; Hamilton, Hall \& Costa 1985) and Crab (Rankin
et al. 1988) pulsars. These variations are probably caused by a magnetized
cloud moving into or out of the line of sight (Hamilton et al. 1985).

Recently, Rand \& Lyne (1994) and Qiao et al. (1995) also published a number
of new pulsar RMs. Manchester \& Johnston (1995) and Navarro et al. (1997)
report the RMs of PSR B1259-63 and PSR J0437$-$4715, respectively.
Combining our new measurements with all previously published RMs, we have a
total of 318 pulsar RMs.

\begin{figure*}		
\psfig{file=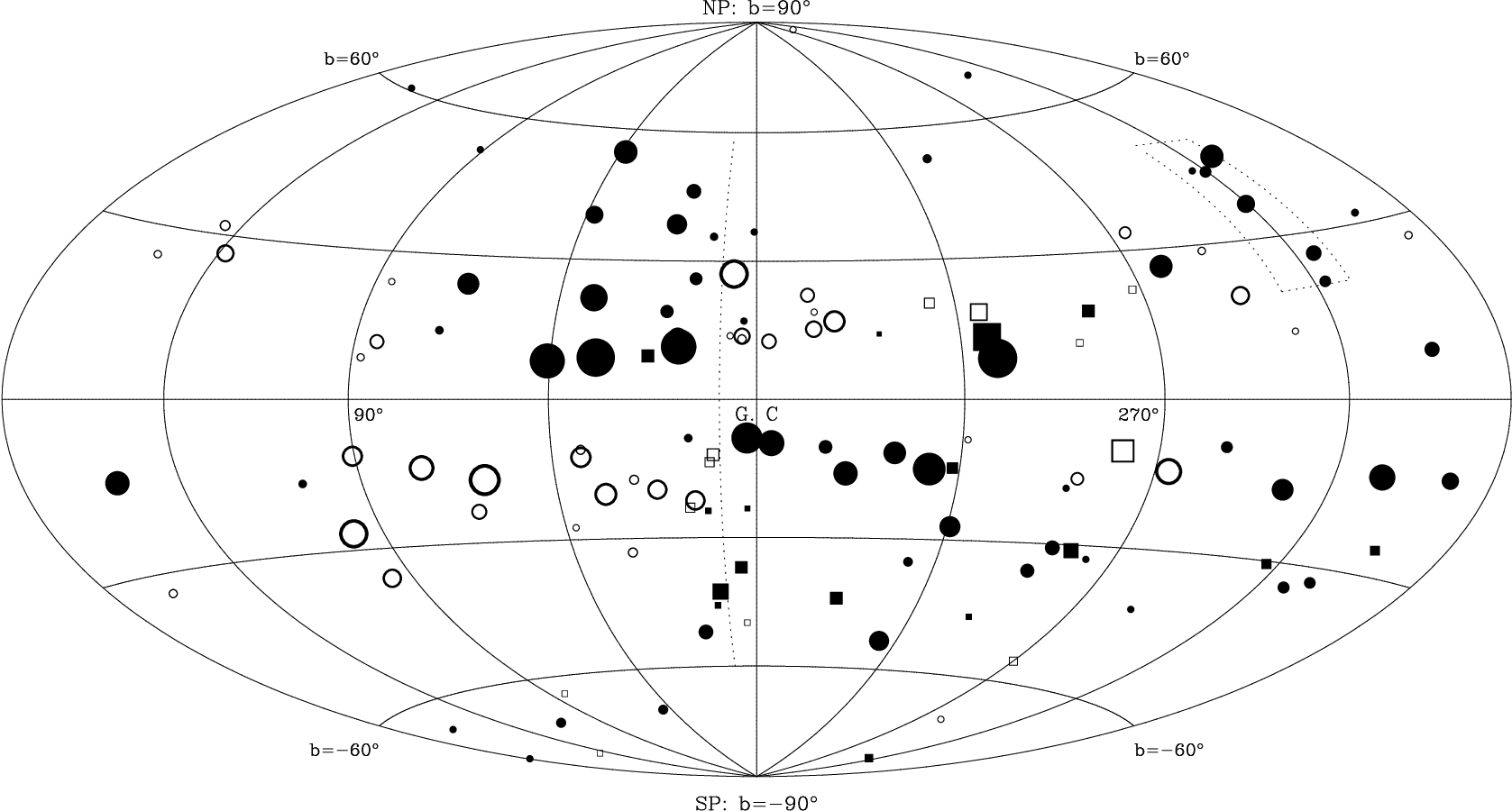,height=78mm,angle=270}
\caption[]{{\bf (a).}Galactic distribution of RMs for pulsars with
$|b|>8\degr$ from Taylor, Manchester \& Lyne (1993), Rand \& Lyne (1994),
Qiao et al. (1995), Navarro et al. (1997) and this work.  The circles
represent previously published RMs, and squares indicate RMs
from this work. Filled symbols represent positive RMs and the area of
the symbols is proportional to $|{\rm RM}|$ within limits of 5 and 150
rad m$^{-2}$. The dotted line indicates the antisymmetry with respect
to $l=8\degr$. The area around $l \sim225\degr$ and $b\sim30\degr$
outlined by dotted lines is discussed in Sect. 3.3.
}
\end{figure*}\addtocounter{figure}{-1}
\begin{figure*}		
\psfig{file=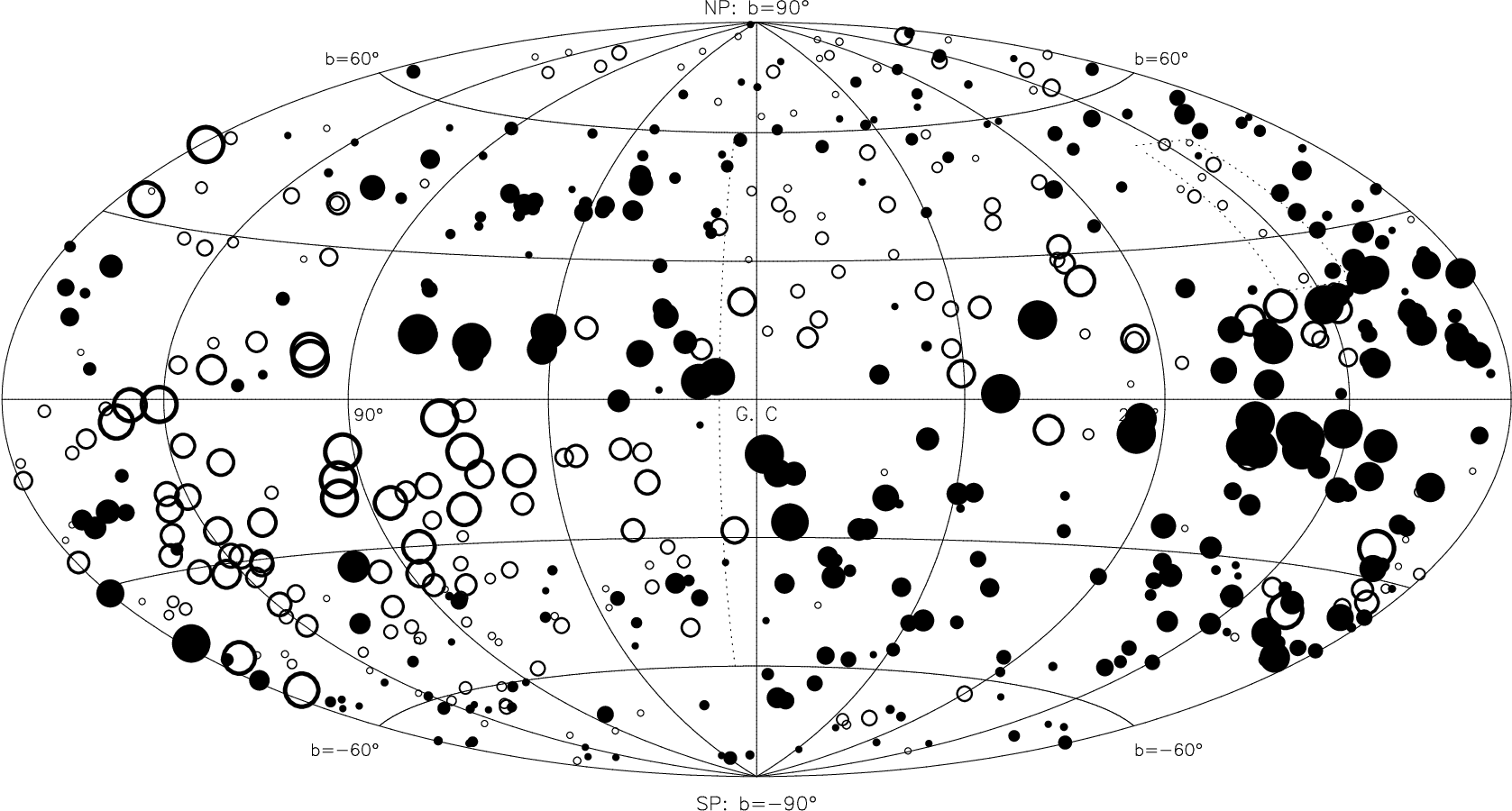,height=78mm,angle=270}
\caption[]{{\bf (b).} Galactic distribution of RMs of selected extragalactic
radio sources (EGRS) after Han et al. (1997). Symbols are plotted as in
Fig. 2(a).  }
\end{figure*}

\section{Magnetic field features at high Galactic latitudes}
\subsection{The antisymmetric RM sky}
In Fig.~2(a), we show the Galactic distribution of RMs of all pulsars with
known RMs and at latitudes $|b|>8\degr$. The latitude selection is made to
emphasize the magnetic field in the thick disc or halo.  Fig.~2(b) shows the
filtered RM distribution of extragalactic radio sources (EGRS) of Han et
al. (1997), in which RMs discrepant from their nearby neighbours by more
than three standard deviations are omitted to more clearly show the
large-scale structure of the Galactic field. The antisymmetric distribution
is clearly evident in the inner Galactic quadrants. With a few exceptions,
the new pulsar RM data shown in Fig.~2(a) are consistent with this
antisymmetric RM distribution. Separating the pulsar RM data into three
groups at different distances ($<2$, 2 -- 4, and $>4$ kpc) shows that the
antisymmetry persists, and nearby pulsars show it at higher latitudes. These
facts, together with the antisymmetry seen in the RM distribution of EGRS,
suggests that the azimuthal fields with opposite directions above and below
the Galactic plane are pervasive in the thick disc or halo of our
Galaxy. One discrepant region, seen in both pulsar and EGRS RMs, is near
$l\sim310\degr$, $b\sim10\degr$, where positive RMs are found.  This region
is discussed in Sect. 4.4.

With a few exceptions, both pulsar and EGRS RM data in the region
$0\degr<l<8\degr$ have the same sign as those in the fourth Galactic
quadrant, $l>270\degr$. These sources extend over both positive and negative
latitudes and over a range of distances, suggesting a large-scale
effect. This result may just reflect field irregularities in this region,
but if confirmed to be large-scale, it would suggest that the magnetic field
in the halo has a spiral component. If so, the indicated pitch angle
is $+8\degr$, {\it opposite} in sign to that of fields in the Galactic
disc. More RMs of distant pulsars and EGRS around this Galactic longitude
are needed to verify this intriguing result.

\begin{figure}  
\psfig{file=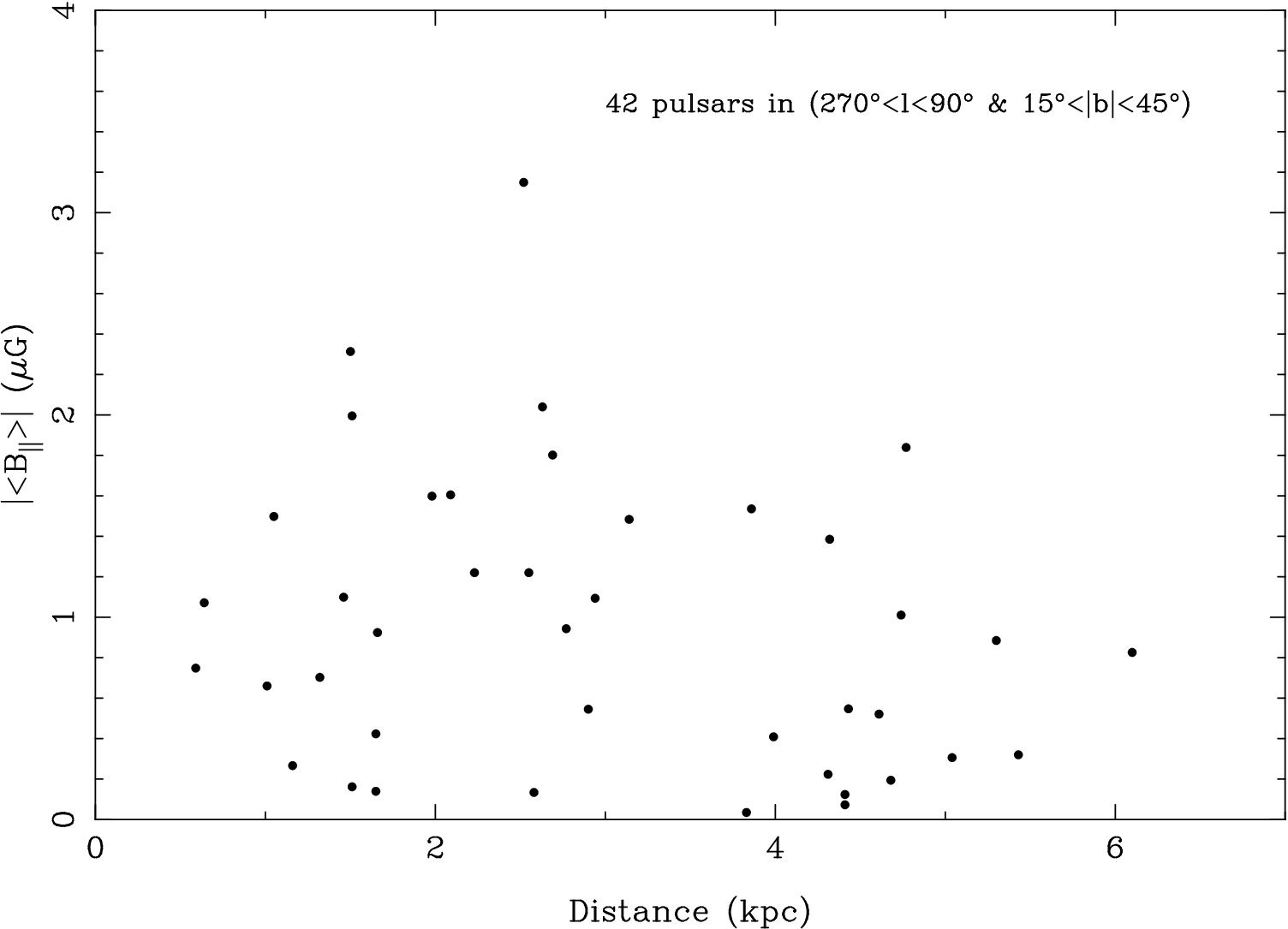,height=55mm,width=75mm,angle=270}
\caption{Average field strength along the line of sight versus distance for
pulsars with $270\degr<l<90\degr$ and $15\degr<|b|<45\degr$.}
\end{figure}

To estimate the strength of the large-scale magnetic field in the inner
quadrants of the Galactic halo, we plot in Fig.~3 the average field strength
along the line of sight versus distance for pulsars with
$15\degr<|b|<45\degr$. Allowing for the effects of irregular fields and
curvature of the field on scales larger than $2 - 3$ kpc, this figure
suggests that the large-scale field in the halo has a strength of $1-2
\;\mu$G.

\subsection{A radial field reversal around $\bmath{l=225\degr}$,
$\bmath{20\degr < b < 50\degr}$ ?}

In the region around $l=225\degr$ and $20\degr < b < 50\degr$, outlined in
Fig. 2(a) and Fig. 2(b), pulsar RMs are positive whereas the filtered EGRS
RMs are negative. The pulsars in this region are mostly at distances of
$1\sim 2$ kpc, and so the mean field over this path length is directed
toward us. However, the mean field through the entire Galactic halo in this
direction is directed away from us on average.  It appears that there is a
reversal of (at least the radial component of) the field at a distance of at
least a few kpc. Although more RMs in this region are desirable to confirm
this reversal, it demonstrates the large extent of the magnetoionic halo of
our Galaxy and possible field structure within it. Recently, Haffner et
al. (1998) have shown that very long H$_{\alpha}$ filaments exist in this
region. If these features are associated, it would imply that the filaments
are magnetized.

\begin{figure*} 
\psfig{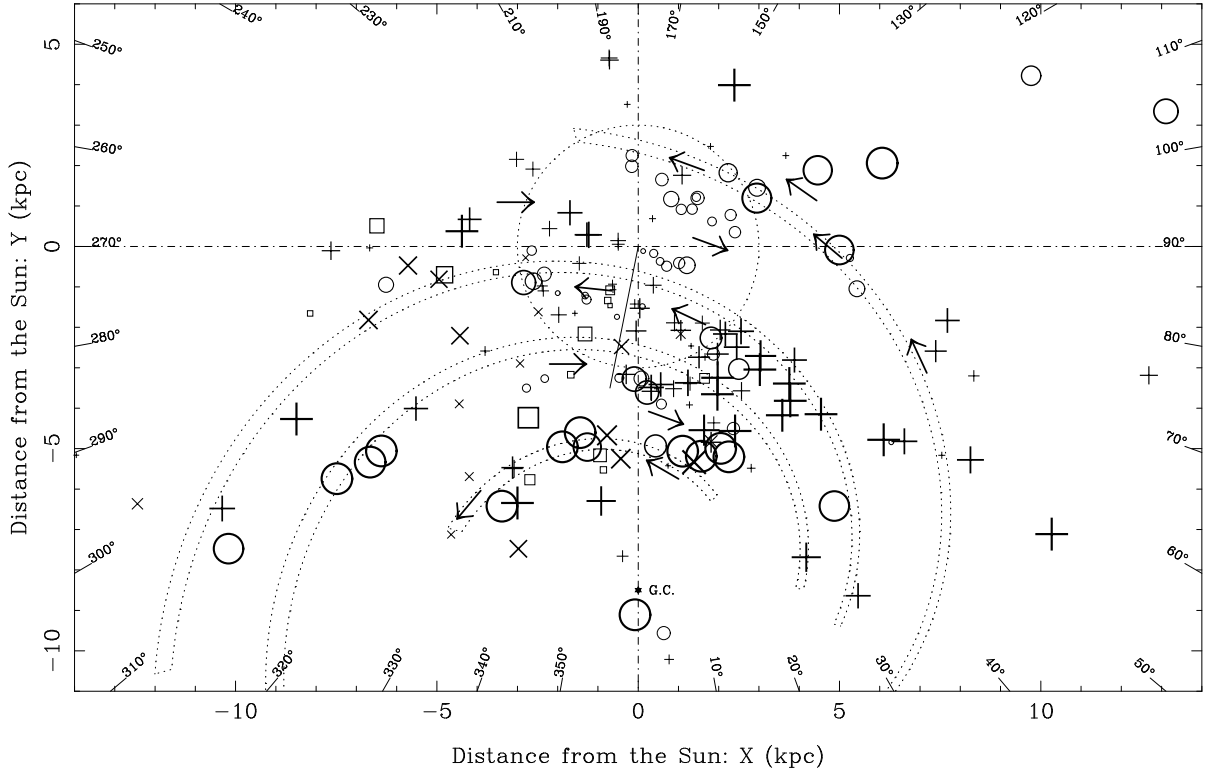}
\caption{The RM distribution of pulsars with $|b|<8\degr$, projected on the
Galactic Plane. The size of the symbols is proportional to the square root
of the RM, with limits of 5 and 250 rad m$^{-2}$ with positive RMs
represented by a cross and negative RMs by a circle. New measurements are
indicated by $\times$ and open squares. The directions of the prevailing
Galactic magnetic field, as revealed by pulsar RMs, are indicated by
arrows. The approximate locations of four spiral arms are indicated; from
the centre out they are, respectively, the Norma arm, the Crux-Scutum arm,
the Carina-Sagittarius arm and the Perseus arm (Georgelin \& Georgelin 1976;
Downes et al. 1980; Caswell \& Haynes 1987). The short line at $l=352\degr$
separates RMs of opposite sign as discussed in the text, and the dotted
circle has a radius of 3 kpc.}
\end{figure*}

\subsection{Vertical component of local magnetic field}

The strength of the magnetic field component perpendicular to the Galactic
Plane is a very important parameter since it affects the maximum particle
energy in the Galactic halo and hence theories for the origin of cosmic rays
(Biermann 1997). It is also an important indicator of field parity in dynamo
models (Zweibel \& Heiles 1997). Han \& Qiao (1994) used the RMs of EGRS at
the two Galactic pole regions to investigate this `vertical' component of
the local magnetic field. They concluded that the local vertical field $B_V$
has a strength of 0.2 -- 0.3 $\mu$G and is directed toward the north
Galactic pole. Because of its importance and the difficulty in measuring it
in other galaxies, we redetermine $B_V$ based on the filtered RM data for
EGRS (Han et al. 1997) and on pulsar RMs in the Galactic pole regions.

\begin{table}  
\caption{Parameters of pulsars in the south polar region}
\setlength{\tabcolsep}{1.7mm}
\begin{tabular}{ccccrr}
\hline 
 PSR J  &  $l$  &   $b$ &  Dist  & \multicolumn{1}{c}{RM} & 
 \multicolumn{1}{c}{$\langle B_{||} \rangle$ }\\
  & ($\degr$) & ($\degr$) & (kpc) & \multicolumn{1}{c}{(rad m$^{-2}$)}
  & \multicolumn{1}{c}{ ($\mu$G)}\\
\hline
0108$-$1431& 140.9& $-$76.8& 0.10&$-$1$\pm$3~!&$-$0.7$\pm$2.0 \\
2330$-$2005& !49.4& $-$70.2& 0.49& 9.5$\pm$0.2&   1.4$\pm$0.3 \\
0034$-$0721& 110.4& $-$69.8& 0.68& 9.8$\pm$0.2&   1.1$\pm$0.1 \\
0152$-$1637& 179.3& $-$72.5& 0.79&   2$\pm$1~!&   0.2$\pm$0.1 \\
0206$-$4028& 258.6& $-$69.6& 0.88&$-$4$\pm$5~!&$-$0.4$\pm$0.5 \\
0134$-$2937& 230.2& $-$80.2& 1.78&  13$\pm$2~!&   0.7$\pm$0.1 \\
0151$-$0635& 160.4& $-$65.0& 1.93&   2$\pm$4~!&   0.1$\pm$0.2 \\
2346$-$0609& !83.9& $-$64.1& 1.96&$-$5$\pm$1~!&$-$0.3$\pm$0.1 \\
\hline
\end{tabular}
\end{table}

We take sources with $|b|>60\degr$ and compute the mean vertical component
of the field assuming the Taylor \& Cordes (1993) model for the Galactic
electron distribution. To eliminate any possible contamination by the fields
of the North Polar Spur, we do not use RMs of EGRS in the region of $
60\degr<b<80\degr$ and $40\degr<l<280\degr$, where the North Polar Spur
(Loop I: Berkhuijsen et al. 1971) goes over the polar region (Sofue 1994). A
model fit (cf. Sect. 4.3 of Han \& Qiao 1994) to $\langle B_{||}\rangle$
values for the remaining 53 EGRS gives a value for the vertical component of
$B_V = - 0.37 \pm 0.14\;\mu$G. The minus sign means that the vertical field
is directed toward the north Galactic pole. The RM contribution of this
component is about 3 rad~m$^{-2}$ toward the southern pole and $-3$
rad~m$^{-2}$ toward the northern pole. If we subtract this contribution, the
RMs of EGRS in the polar regions, mainly intrinsic or from the intergalactic
medium, have a variance of 11 rad~m$^{-2}$.

There are few pulsar RMs in the north polar region, but we can use the RMs
of eight pulsars in the south polar region ($b<-60\degr$) as an independent
check of our result. In Table 2, we list the parameters of these pulsars,
sorted by distance. We see that four of the five well-determined RMs (error
less then $|RM|$) are positive, which is qualitatively consistent with our
conclusion from the EGRS.

\section{Magnetic Features in the Galactic disc}

Fig. 4 shows the RM distribution of 189 low-latitude pulsars ($|b|<8\degr$),
projected onto the Galactic Plane. Separately plotting the data for pulsars
above and below the Galactic plane shows little significant difference. In
this section, we consider the structure of the local field using RMs of
pulsars within 3 kpc of the Sun, and then discuss the evidence for reversals
in the field at greater distances.

\subsection{The local field}
Most previous analyses of pulsar RMs (e.g. Manchester 1974; Lyne \& Smith
1989; Rand \& Kulkarni 1989) have concluded that the local field is directed
toward $l\approx95\degr$, i.e., with a pitch angle of $\sim5\degr$.  This is
not consistent with spiral models in which the field lines have a negative
pitch angle similar to that of the spiral arms. This discrepancy may be
partly explained by selection effects (cf. Han \& Qiao 1994), especially the
lack of pulsar observations in the fourth (southern) quadrant, and by the
influence of the large negative RMs at $l\sim 110\degr$ to $120\degr$. This
group of large negative RMs may be related to the stellar complex in the
Perseus arm at these longitudes (e.g. Fig. 2 of Efremov 1995). Passage
through a dense ionized region could result in both the large DMs and the
large RMs of these pulsars; the distances of these pulsars may be
over-estimated.

All pulsars with distances less than 3 kpc in the region
$352\degr<l<70\degr$ (indicated by a short line in Fig. 4) have positive
RMs, in contrast to negative RMs of $270\degr<350\degr$, consistent with a
pitch angle for the local field of $\sim -8\degr$. Near and outside the
solar circle, the nearby pulsars indicate a regular field running in the
clockwise direction whereas those closer to the Galactic Centre indicate a
counter-clockwise field.

\subsection{Field reversals in the Perseus arm and beyond}

Evidence for field reversals in the outer Galaxy (i.e. in or about the
Perseus arm) was previously found by Lyne \& Smith (1989) and Clegg et
al. (1992).  Lyne \& Smith (1989) noticed that in the region of
$105\degr<l<135\degr$ the RM values of extragalactic radio sources are
notably smaller than those for most distant pulsars and that for some the
signs are reversed, indicating a field reversal beyond the region occupied
by the pulsars. Clegg et al. (1992) concluded from the RMs of extragalactic
radio sources that there may be a field reversal exterior to the Solar
Circle.
\begin{figure}
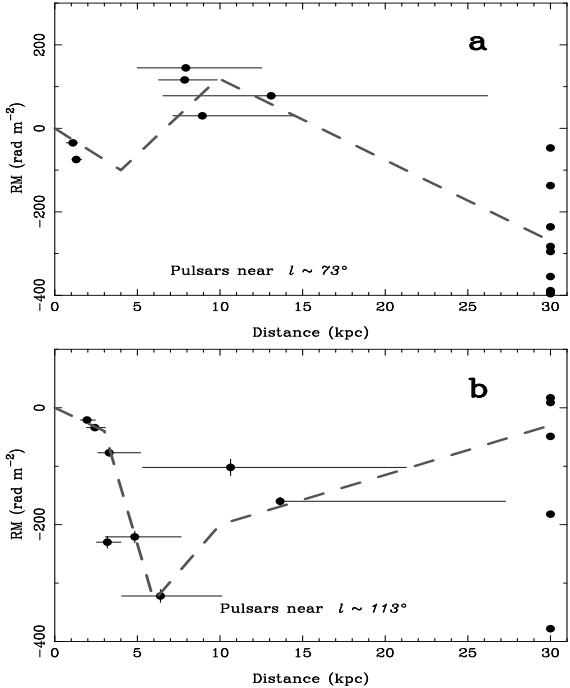
	
\begin{tabular}{c}
\psfig{file=l_73rm_d.ps,height=45mm,width=7.5cm,angle=270}\\
\psfig{file=l_110.ps,height=45mm,width=7.5cm,angle=270}
\end{tabular}
\caption{RMs of pulsars and extragalactic radio sources {\bf (a).} for $l =
73\degr \pm5 \degr$, and {\bf (b).} for $l=113\degr \pm
6\degr$. Uncertainties in the RMs and distances are indicated for each
pulsar.  Extragalactic radio sources are plotted at 30 kpc. General trends
in the RM variations with distance are indicated by dashed lines.}
\end{figure}

As shown in Fig. 4, all distant pulsars, i.e. those outside the Perseus arm,
in directions between $l=55\degr$ and $80\degr$, have positive RMs.  This
contrasts with the negative RMs of nearby pulsars in these directions.
Following Clegg et al. (1992) and Rand \& Lyne (1994), we plot in Fig. 5(a) RM
versus distance for low-latitude pulsars and extragalactic radio sources
within $5\degr$ of $l=73\degr$. The RM increase from negative values of
nearby pulsars to the positive ones of these distant pulsars clearly
indicates a reversal in the mean field direction in or around the Perseus
arm, occuring at a distance $\sim 5$ kpc. However, the RMs of the EGRS are
again negative, indicating a further reversal at distances $\ga 15$
kpc. Clearly, the magnetoionic disc of our Galaxy is very extended (cf. Han
\& Qiao 1994; Clegg et al. 1992).

RMs in directions around $l\sim113\degr$ also give some indication of a
field reversal at a distance of about 5 -- 7 kpc (but note that the pulsar
distances may be overestimated as mentioned in Sect. 4.1). As shown
in Fig. 5(b), the RMs of pulsars and EGRS beyond this distance become less
negative in general, indicating a positive contribution to the RM from
the reversed field in the region beyond the Perseus arm.

Therefore, both of these figures indicate a counter-clockwise field in the
region in and beyond the Perseus arm (viewed from the north), in contrast to
the clockwise field between the Perseus and Sagittarius arms. Positive RMs
around $l\sim150\degr$ are consistent with this reversed field. However, in
the direction of $180\degr<l<190\degr$, RMs are scattered both positive and
negative. In this direction, the lines of sight are approximately
perpendicular to the large-scale field, allowing smaller scale
irregularities to dominate.

\begin{figure}  
\psfig{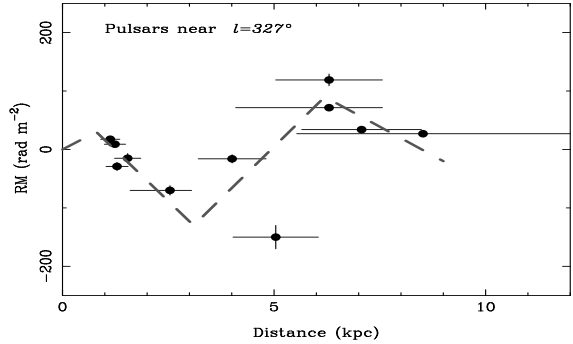}
\caption{ Variation of pulsar RMs with distance in the direction $l =
327\degr \pm5\degr$. Uncertainties are indicated. Trends in the RM
variations are indicated by the dashed lines.}
\end{figure}
\subsection{Evidence for a third field reversal inside the solar circle?}

Previous work has identified two field reversals inside the Solar Circle, in
the Carina-Sagittartus arm and in the Crux-Scutum arm (e.g. Vall\'ee 1991;
Rand \& Kulkarni 1989; Rand \& Lyne 1994).  A third field reversal in the
Norma arm is predicted by the BSS models of Sofue \& Fujimoto (1983) and Han
\& Qiao (1994).  Considering the spiral form of the Norma arm, we choose 
directions around $l\sim 327\degr$ and $l\sim10\degr$ to examine the RM
variations with distance. If the magnetic field follows the spiral arms, as
has been seen in nearby spiral galaxies (Beck et al. 1996; Kronberg 1994),
it would be almost parallel to the line of sight and hence have the largest
RM contribution in these directions.

For $l\sim 327\degr$ (Fig. 6), the local field within 1 kpc is responsible
for positive RMs of the nearby pulsars, indicating a clockwise field near
the Sun. Between $\sim 1$ and $\sim 4$ kpc, between the Carina-Sagittarius
arm and the Crux-Scutum arms, the field direction is reversed. Beyond this
distance, the RMs show an increasing trend, indicating a second reversal to
a clockwise field between the Crux-Scutum and Norma arms.  Beyond this, at
distances $\ga6$ kpc, there is some indication that the RMs are decreasing
again, suggesting a possible third reversal in or near the Norma arm.

\begin{figure*}	
\psfig{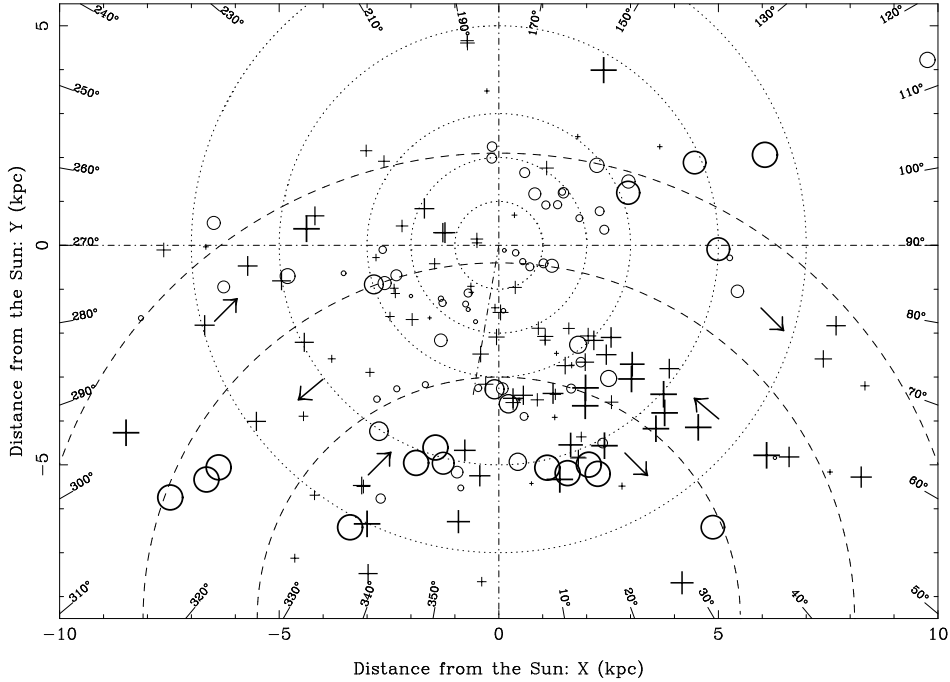}
\caption{Distribution of pulsar RMs and the concentric-ring model sketched
according to Rand \& Kulkarni (1989) and Rand \& Lyne (1994).  Field
directions expected for this model are indicated by arrows and the dashed
circles show where the field lines reverse. The dotted circles indicate
equal distances from the Sun.}
\end{figure*}
\begin{figure*}	
\psfig{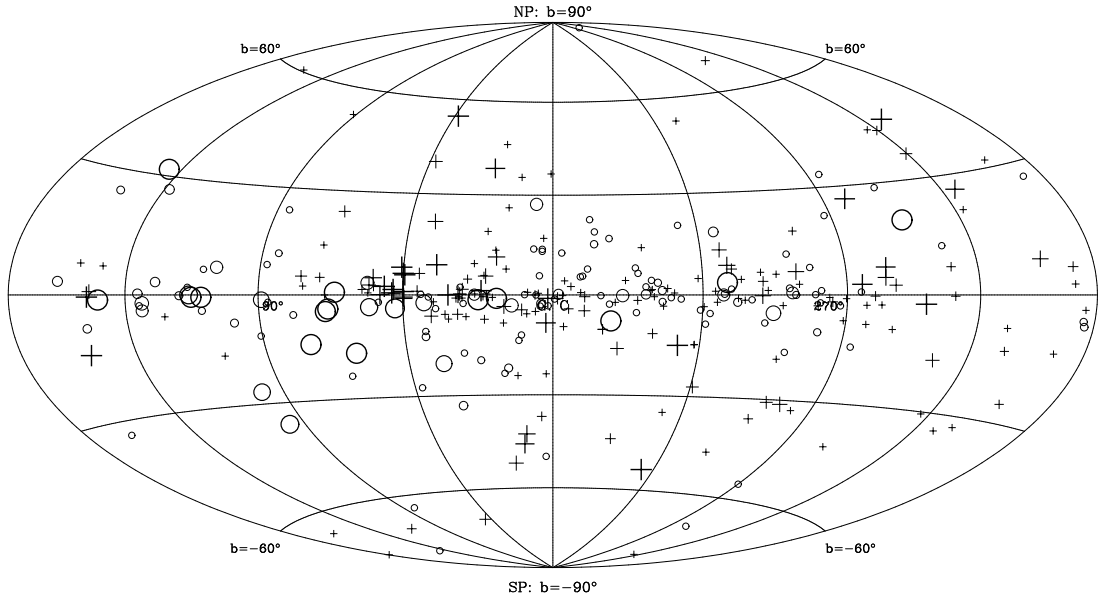}
\caption{Distribution of $\langle B_{||}\rangle$ of all measured pulsar RMs
in Galactic coordinates. Plus signs indicate the average field is directed
towards us, and circles indicate the average field is directed away from
us. The size of symbols is proportional to the field strength within limits
of 0.8$\mu$G and 2.5$\mu$G. }
\end{figure*}

\subsection{Positive RMs in the fourth Galactic quadrant}

At longitudes between $280\degr$ and $310\degr$ and distances greater
than 3 kpc, there is a group of pulsars with positive RMs (Fig. 4).
These pulsars lie between the Carina and Crux arms and, according to
the BSS model, should have negative RMs to complement the positive RMs
at longitudes between $35\degr$ and $55\degr$.  As shown in Fig. 7,
these RMs are consistent with the concentric model, but the many
positive RMs in the range 300 to $320\degr$ at distances greater
than 2.5 kpc are not.

Fig. 8 shows that the positive RMs and hence positive $\langle B_{||}
\rangle$ at low latitudes in the region between $280\degr$ and $310\degr$
are mostly above the Galactic Plane, whereas those with negative $\langle
B_{||}\rangle$ mostly lie below the Plane. This reversal from the general
pattern is probably related to an anomalous region in this direction at
distances greater than about 3 kpc. It may represent a large-scale component
of the random field or a systematic distortion of the uniform field, and
should be investigated in future with more pulsar RM data. Neither Loop I
nor the Gum Nebula can account for it as both of these are much closer. 

\section{Discussion}

The results presented in Sections 3 and 4 show that the magnetic field in
our Galaxy has a very complicated structure. In the thick disc or halo,
there is an azimuthal or spiral field with odd symmetry with respect to
Galactic plane. The vertical field in the nearby halo is another
discriminator for odd parity (Zweibel \& Heiles 1997). There is some
evidence for such a vertical field in both the pulsar and EGRS data, but at
present its statistical significance is weak. In the disc, the field
structure is probably related to the spiral arms, as is seen in nearby
galaxies (see Beck et al. 1996). Discrimination between the three main
models for the disc field proposed so far, the concentric-ring model, the
axisymmetric spiral (ASS) model, and the bisymmetric spiral (BSS) model,
relies mainly on estimates of the pitch angle of the large-scale field
component and the presence and number of field reversals as a function of
Galactocentric radius. Unfortunately, the dominance of irregular structures
in the field (Ohno \& Shibata 1993) makes discrimination between the
various models difficult.

Only the concentric-ring model has a pitch angle of zero; the two spiral
models predict a pitch angle close to that of the spiral arms. In our
Galaxy, both starlight polarization (Heiles 1996a) and the Faraday rotation
measurements (Han \& Qiao 1993, 1994; Sect. 4 above) suggest a pitch angle
for the local field of $\sim -8\degr$, somewhat less than the pitch angle of
spiral arms which is between $-8\degr$ and $-14\degr$, with a best estimate
of $-12\degr$ (Vall\'ee 1995). As discussed in Sect. 4.4, many of the
observed RMs, in particular, those around $l\sim 310\degr$, are not
consistent with the ring model. Similarly, the group of positive
RMs at large distances around $l\sim70\degr$ are inconsistent with this
model. Furthermore, there are theoretical difficulties with a circular field
structure (Beck et al. 1996).  We therefore consider that there is little
support for the concentric-ring model of the Galactic magnetic field.

In its pure form, an ASS field has no reversals as a function of
Galactocentric radius. Such a field structure can be generated by dynamo
action. However, in one-dimensional simulations, Poezd et al. (1993) showed
that the ASS field generated by dynamo models could have field
reversals. The form and number of these is dependent on assumptions about
the initial conditions and the number of galactic rotations and these results
need to be confirmed with more realistic (3D) simulations. In external
galaxies, field reversals associated with spiral arms have been observed only
in BSS candidates (e.g. Beck et al. 1996; Zweibel \& Heiles 1997).

In our Galaxy, there is good evidence for one and weaker evidence for
two reversals in and beyond the Perseus arm (see also Heiles 1995, 1996ab).
Inside the solar Circle, two reversals seem to be more or less certain and
we give marginal evidence for a possible third reversal associated with
the Norma arm. This means that in all there are probably three, maybe
up to five, reversals in the direction of the large-scale Galactic field
as a function of Galactocentric radius. This number of reversals is less
consistent with ASS models, which have no intrinsic reversals, than with BSS
models which have reversals associated with the spiral arms.

On the other hand, the observed antisymmetry of high-latitude RMs suggests
that the field in the thick disc and halo of the Galaxy has an A0 form and
hence probably generated or maintained by a large-scale dynamo (Han et
al. 1997). Yet there is good evidence for close connections between the disc
and halo field. For example, around $l\sim 45\degr$, there are positive RMs
extending from $b\sim 0\degr$ to high positive latitudes (Fig.~8). At high
latitudes, these positive RMs form part of the antisymmetric halo field.
Many of the disc pulsars
with positive RMs are at distances of several kpc (Fig.~4), and their
existence is a large part of the argument for the field reversal inside the
Sagittarius arm.

Clearly, the question of the origin and structure of the Galactic magnetic
field has not yet been answered. There is evidence for ordered magnetic
fields in early epochs of galaxy formation (Udomparsert et al. 1997; Athreya
et al. 1998), but dynamo processes (Parker 1997; Kulsrud 1997) will very
likely amplify and modify the field structure. More realistic simulations
(e.g. Bykov et al. 1997; Rohde \& Elstner 1998) show wide variations in the
form of the large scale field. Furthermore, the field structure can be
complicated by spiral streaming (Lou \& Fan 1998; Moss 1998), stellar wind
bubbles and supernova events. It is not surprising that the observed field
structure in galactic disks cannot yet be fully explained by dynamo models
(see the reviews by Kronberg 1994 and Beck et al. 1996).

\section{Conclusions}

In this paper, we present RMs for 63 pulsars, 54 of which have no previously
published RM data. RMs of three pulsars are shown to have significant
variations over timescales of several years. These results reveal the
complicated structure of the Galactic magnetic field.

At high Galactic latitudes, the results are largely consistent with the
antisymmetric field pattern found by Han et al. (1997). Marginal evidence
was found for a leading spiral component in the halo field. An interesting
region of reversed radial field at distances of a few kpc was found in the
region $l\sim 225\degr, 20\degr <b< 50\degr$.

Previous observations of fields in the Galactic disc have provided good
evidence for two reversals in the large-scale azimuthal field, one inside
the Carina-Sagittarius arm at a Galactocentric radius $R\sim 8$ kpc
($R_{\sun} = 8.5$ kpc), and the other near the Crux-Scutum arm at $R\sim
5.5$ kpc, and weaker evidence for a reversal near or beyond the Perseus arm
at $R\ga 10$ kpc (see Heiles 1995, 1996b). The new results strengthen the
evidence for the Perseus
arm reversal and provide marginal evidence for two additional reversals, one
beyond the Perseus arm and one in the inner Galaxy near the Norma arm at
$R\sim 3$ kpc. These results, coupled with good evidence that the
large-scale disc fields are spiral with a pitch angle $\sim -8\degr$, are
most consistent with a bisymmetric spiral (BSS) form for the disc
field. However, positive RMs observed for pulsars located between the
Carina-Sagittartus and Crux-Scutum arms in the fourth Galactic quadrant
cannot be explained by any spiral model. There is some evidence that the
disc and halo fields are connected, further complicating their
interpretation.

\section*{Acknowledgments}

We thank Drs  Elly M. Berkhuijsen, Rainer Beck and the referees for their
careful reading of the manuscript and their valuable comments.
JLH thanks the Su Shu Huang Astrophysics Research Foundation of CAS and the
Director Foundation of BAO for financial support for travel to the ATNF. He
also acknowledges the hospitality from Prof. R. Wielebinski during his stay
at Max-Planck-Institut f\"ur Radioastronomie
and financial support from the National Natural Science Foundation (NNSF)
of China, the Research Foundation from the Astronomical Committee of CAS.
GJQ thanks the NNSF of China for support for his visits at ATNF and
acknowledges the finacial support from the Climbing Project -- the National
Key Project for Fundamental Research of China.
JLH and GJQ also thank the ATNF for its hospitality and acknowledge support 
from
the Bilateral Science and Technology Program of the Australia Department of
Industry, Science and Tourism.
The Parkes telescope is part of the Australia Telescope which is funded by
the Commonwealth of Australia for operation as a National Facility by CSIRO.

\end{document}